# SmartState: An Automated Research Protocol Adherence System

**Samuel E. Armstrong, MS[1], Mitchell A. Klusty, BS[1], Aaron D. Mullen, BS[1], Jeffery C. Talbert, PhD, FAMIA[1], Cody Bumgardner, PhD[1]**
**[1]Institute for Biomedical Informatics, University of Kentucky, Lexington, KY**

**Abstract**
*Developing and enforcing study protocols is crucial in medical research, especially as interactions with participants become more intricate. Traditional rules-based systems struggle to provide the automation and flexibility required for real-time, personalized data collection. We introduce SmartState, a state-based system designed to act as a personal agent for each participant, continuously managing and tracking their unique interactions. Unlike traditional reporting systems, SmartState enables real-time, automated data collection with minimal oversight. By integrating large language models to distill conversations into structured data, SmartState reduces errors and safeguards data integrity through built-in protocol and participant auditing. We demonstrate its utility in research trials involving time-dependent participant interactions, addressing the increasing need for reliable automation in complex clinical studies.*

## 1. Introduction

Improving the efficiency, quality, and impact of research studies is a key goal of the National Center for Advancing Translational Sciences (NCATS) Strategic Plan[1]. Using data science approaches to accelerate translation is a central NCATS objective involving expanding informatics and strategies to simplify and automate research processes. We present one example of this strategy focused on protocol adherence. Research studies involving human subjects require regular interactions with participants, including recording study parameters or assessing participant status. Traditionally, these interactions can occur manually by calling, texting, or emailing the participant. However, this becomes increasingly cumbersome for large participant groups, especially when timely interactions are vital to study results. Moreover, ensuring participants stay within set study guidelines becomes increasingly challenging as the number of participants increases. Even with dedicated outreach coordinators, researchers can become overwhelmed, increasing the need for interventions. This increases the likelihood of errors or inconsistencies. As a result, there is a significant need for automated systems that manage these interactions while maintaining accuracy and timing. SmartState is employed in three research studies related to time-restricted eating, optimal medication administration, and plant-based diets. In each of the three studies we will discuss, manual communication methods were used initially, which was acceptable for small cohorts. However, frequently communicating with each participant became tedious as more participants enrolled. Thus, additional time and effort were required to record responses when participants needed help to stay within study parameters.

The benefits of our proposed system apply to both researchers and software developers by simplifying the development and configuration process for a study. By automating the management of participants, communications, and study data, the system overcomes the limitations of traditional rule-based logic, which requires defining explicit rules for every possible participant action. This approach becomes unmanageable as the complexity of study parameters increases. SmartState instead utilizes state-based methods to automatically manage participants, reducing discrepancies and preventing deviations from study protocols. This makes it more efficient and scalable for large studies that can be applied to many research areas.

## 2. Methods

SmartState consists of four components: a messaging service, a state machine, a conversational AI, and a web-based platform. As shown in **Figure 1**, participants engage with the system primarily through a messaging service via texting, emailing, or calling, and can also interact via a chat interface hosted on the web platform. In either case, these interactions are directed to the state machine, which records and initiates interactions and tracks the state of each study participant enrolled in the system. The state machine interprets and processes every incoming message to determine the appropriate next steps.

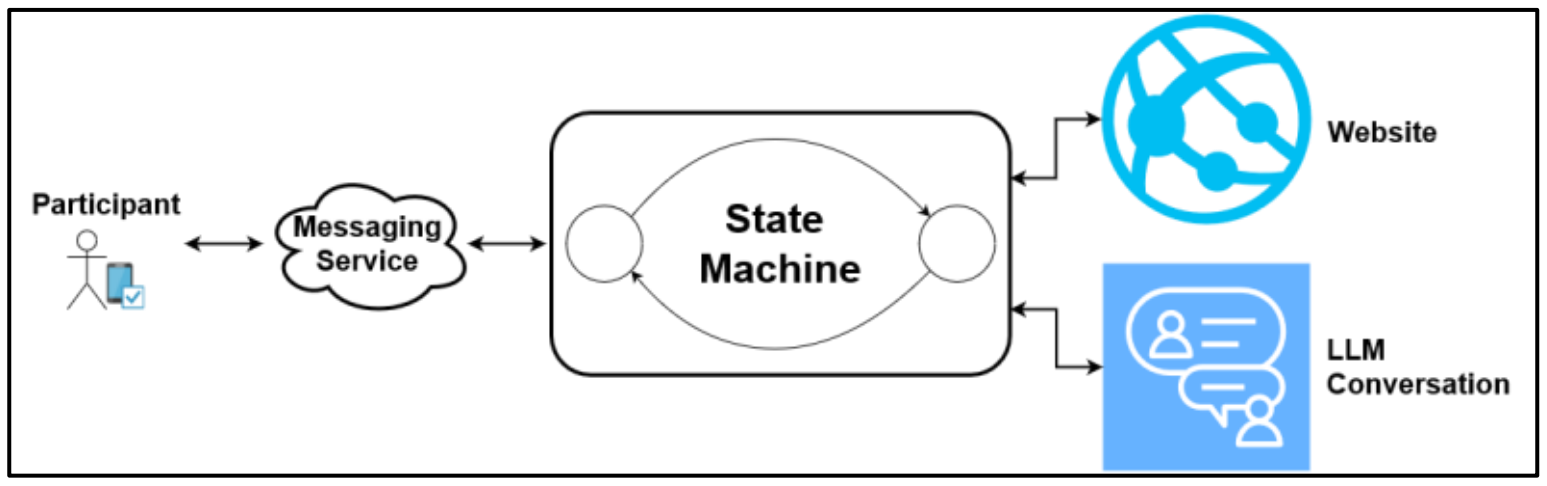

**Figure 1.** Overview of SmartState components and their interactions.

### 2.1 Messaging Service

Depending on a message's content, the system stores the message body (including any attachments), message timestamp, and the participant's phone number. Then, the message is forwarded to the conversational AI for further analysis and response generation. In some instances, the state machine will initiate a new process or send an automated message back to the participant or a study administrator. The components within the system communicate bidirectionally, meaning they can both send and receive information. This two-way communication allows SmartState to handle events dynamically, collect additional information from participants, and promptly notify administrators when specific actions are required.

### 2.2 State Machine

At its core, SmartState uses finite state machines (FSMs) to track participants' progress in a study. FSMs model system behavior as a series of transitions between well-defined states. At each step, specific conditions must be met for a transition to occur. These steps include the completion of a task or the passage of time, ensuring the system moves logically from one state to the next. The states and allowed transitions can be represented as a graph and depicted visually (see **Figure 2**), making the allowed behaviors of the system easier to grasp. Because of this versatility, FSMs are used in many areas, such as circuit design[2], robotics[3], smart homes[4], and autonomous vehicles[5]. Generally, following a path in a graph is straightforward, which facilitates collaboration between developers and clinicians. Creating a graph helps all parties involved in a study in agreement on what actions should happen—as opposed to deciphering lines of code, which can lead to confusion.

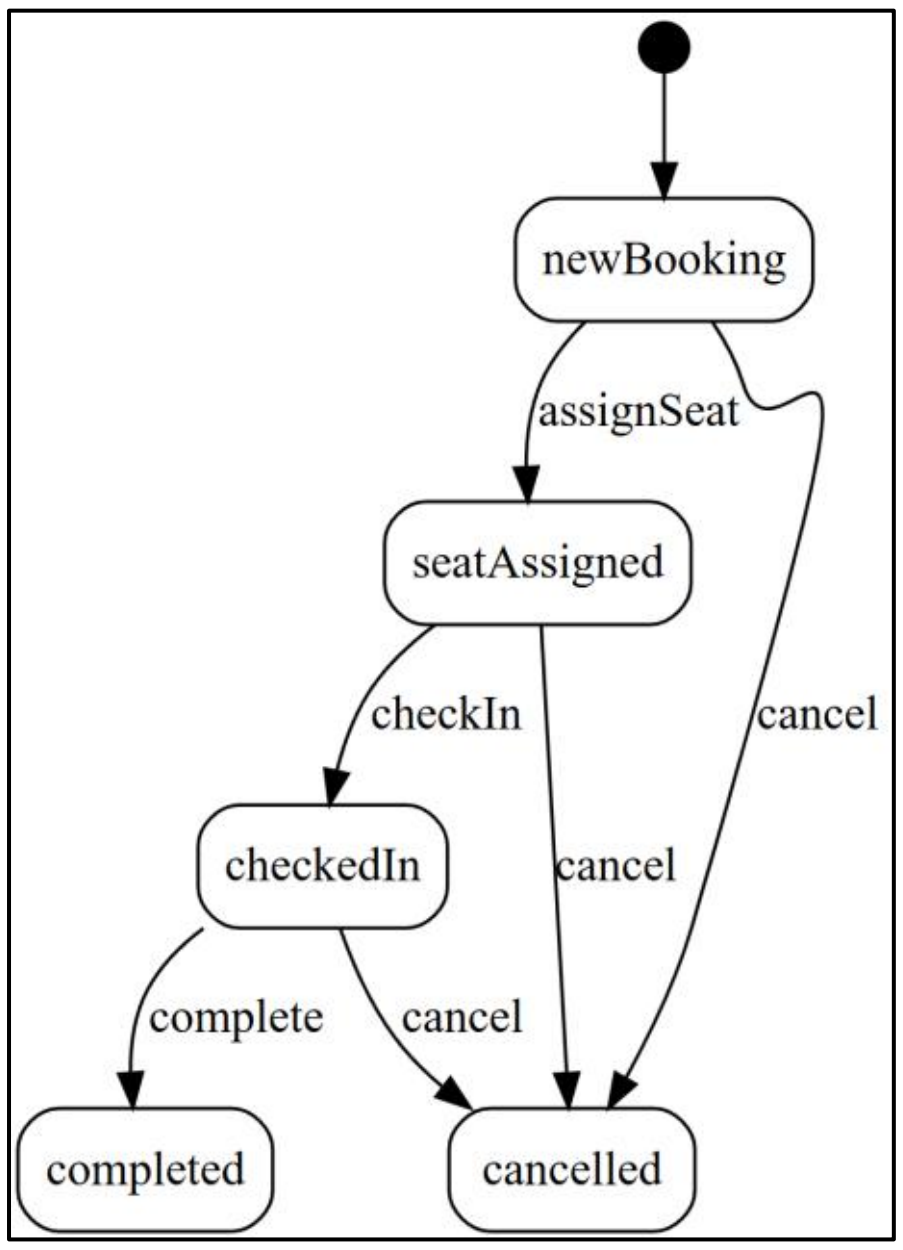

**Figure 2.** An example finite state machine graph for booking an airline seat.

Once a study protocol graph is defined, it must be converted into code that captures the state machine structure and conditions under which transitions may occur. This process is automated, and the generated output is guaranteed correct using a web-based compiler, described in section **3. Architecture**. The compiler can detect errors in graph definitions and generate the required code in a standard form. Java was selected as the output language of this compiler due to its portability and widespread use in the developer community. The generated code is a standard Java class, which can be directly added to SmartState. It can also be extended to add specific functionalities to each state, such as sending messages or querying a database. If additional parameters need to be included, the protocol graph can be updated, and the code can be regenerated with minimal adjustments. Once an FSM is created, the paths defined in the graph are fixed and cannot be deviated from. These fixed paths force participants to pass through specific checkpoints or achieve a specific status before reaching a goal, preventing critical steps from being skipped and maintaining a systematic progression.

FSMs provide the added benefit of capturing detailed logs of every message received or sent, every state entered, and the transition followed. This allows researchers to trace a participant's exact path through a study. Auditing the system in this way prevents a dispute between researchers and participants who claim that they have followed the prescribed orders but, in reality, may have deviated from the required plan.

### 2.3 Conversational AI

Another pillar of SmartState is the interactions between the system and participants. SmartState uses a large language model (LLM) to facilitate this communication. LLMs are advanced artificial intelligence systems designed to understand, generate, and interact with human language in a coherent and contextually relevant manner. These systems are exploding in popularity and use cases, especially in automated participant communications. These models are built upon vast amounts of text data and sophisticated neural network architectures, which can produce human-sounding text responses.

LLMs have their limitations, however. One notable issue is "hallucination," where an LLM generates incorrect, misleading, or fabricated information. Hallucinations can occur because the models may not always accurately interpret or verify the information they retrieve, leading to outputs that can still be incorrect or misleading. Despite advancements, it is challenging to eliminate hallucinations due to the fundamental nature of how LLMs generate responses—based on statistical probabilities rather than factual validation. This intrinsic limitation means that while LLMs can generate impressively coherent text, they are prone to errors, and ongoing efforts are needed to minimize their effect on practical applications.

For this reason, SmartState uses an LLM to better understand natural language inputs from participants, providing structured and thematic data. By using Meta's Llama 3.1 Instruct model[6] as a base, a custom LLM structure was created and accessed through our local LLM self-service provider[7]. Thus, this model runs on a high-performance inference server to retrieve responses quickly without the cost of using a commercial model provider. SmartState can be used with any LLM compatible with OpenAI's programming interface—ChatGPT, Mistral, Llama, and others—by simply adjusting the model's name in the LLM configuration and providing a valid access key from the model provider. As of this writing, Llama 3.1 is Meta's flagship LLM model and is competitive with other commercially available models[8] and free for personal, research, and commercial applications. The Llama model can also invoke external tools programmatically to add additional structure and context to the conversations, allowing it to perform tasks beyond language generation. By integrating with external systems, the LLM can retrieve real-time data, execute specific functions, and automate workflows. For example, when a user responds, the LLM can distill the input text to a set number of arguments that can be passed to a programmed function. Depending on the output of this function, the model is given additional context to reiterate a question or continue a conversation by storing the response and moving to the next question.

An additional FSM is utilized to handle situations where a response is inadequate for a given question. *Adequacy* is defined as the participant being able to provide the necessary content, including clear, concise, and relevant information to satisfy the query without ambiguity or omission. This system operates in conjunction with the response analysis tools described above, which are invoked after each response. The state graph in **Figure 3** illustrates this process. Once provided with the necessary inputs, the function output is used to determine if the system needs to repeat the question to the user with additional context or in an alternate way. Once a user responds satisfactorily, the FSM advances the state to the next question, and this process repeats for each question in a given set.

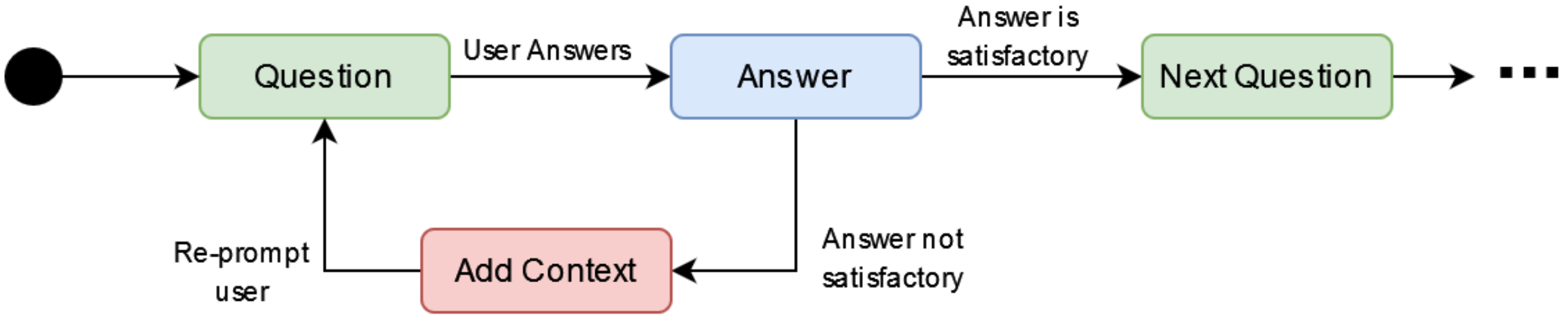

**Figure 3.** State machine graph handling user input. If the response is not satisfactory, add context and ask the question again.

## 2.4 Web Platform

A web-based platform is provided for researchers to access all functionality, participant logs, reports, and messaging utilities (see **Figure 4**). Beyond storing and displaying responses to questions, the web platform facilitates participant registration, communication, and various analysis functions. Authorized researchers can easily access and download participant metrics, enabling seamless interaction with the underlying system. This holistic approach ensures that participants' interactions are captured accurately while researchers benefit from an intuitive, real-time interface that supports data collection and evaluation processes.

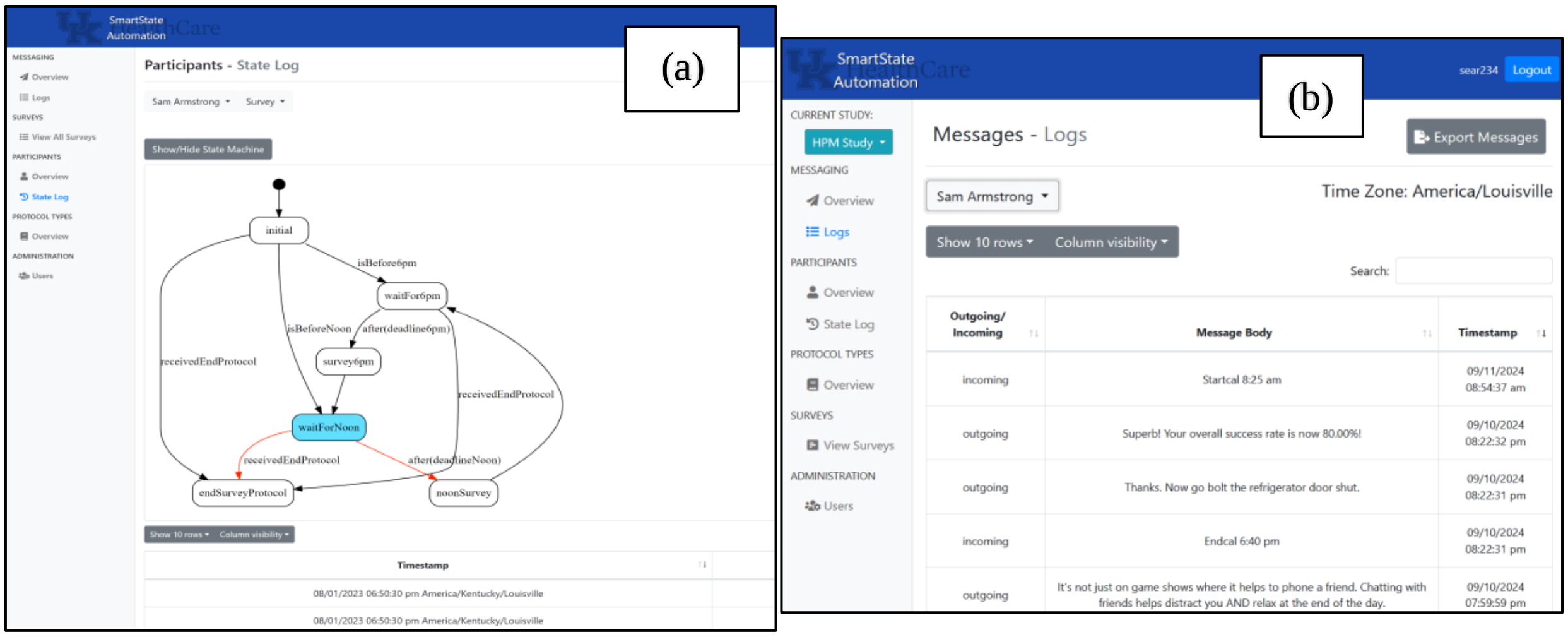

**Figure 4**. **(a)** State machine displayed as a graph with audit logs of states entered and transitions. **(b)** Message logs for a specific participant.

## 3. Architecture

This section provides a more detailed description of the components of SmartState, along with the design decisions that guided the selection of each element. As noted above, finite state machines (FSMs) act as pathways for participants to traverse through a study. The system operates in conjunction with Umple[9], a free online tool that assists with generating FSM code and graphs. Umple is a compiler and modeling tool that uses the unified modeling language (UML) to create graphs and translates them into executable code. Umple is intuitive and offers many examples of UML models to use as starting points. Considering Umple is a compiler, the code generated is guaranteed to be correct and fully compatible with any of the available programming language that Umple supports.

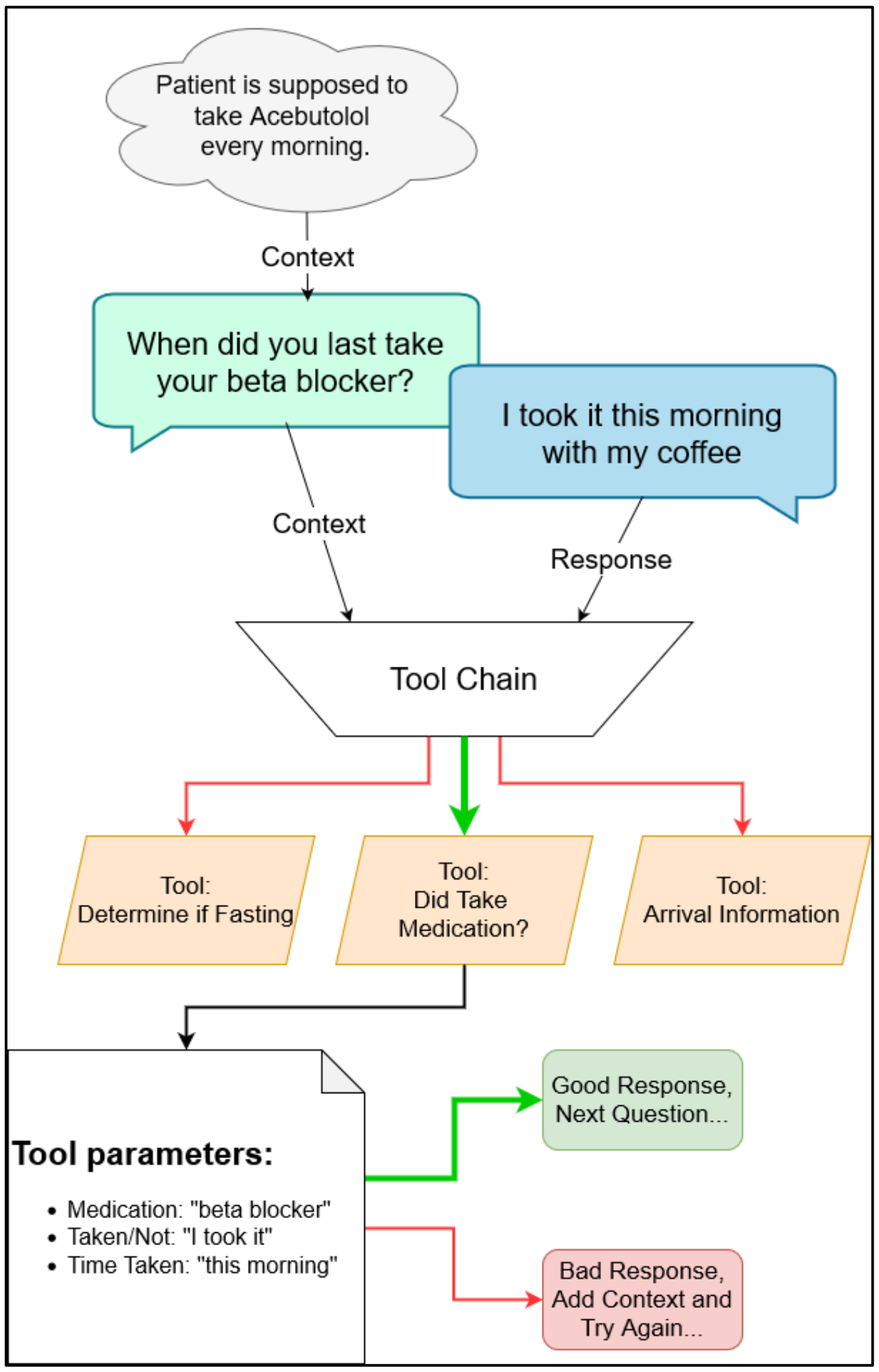

**Figure 5.** Example workflow for a conversation with a participant verifying the use of a beta blocker.

Java code generated by Umple is loaded into SmartState, where it serves as a base class. This class can be extended to add additional functionality, such as database queries, sending messages, or LLM interactions. Extending the class is straightforward, as it follows standard Java rules and can be done either in Java or in Umple before code generation. In practice, we found extending the base class in Java makes modifications to the FSM easier to manage during development. Moreover, each participant is assigned their own instance of the state machine class; an event watcher must be added to instantiate new participants, manage state transitions, and handle external functions. This watcher class, included in SmartState, can be used with both base and extended classes. Once configured, the system will be ready to begin registering and managing participants.

As previously stated, the Llama LLM can use external functions to provide additional context to the conversation. In recent months, Langchain, a framework for developing LLM applications, introduced a method called "tool calling"[10]. Tool calling allows an LLM to extract arguments for a function to check if a response is adequate. Given a set of functions with different input arguments, the LLM will try to match parts of a response to the arguments for each function. The LLM can match one or more functions and automatically execute (or call) these functions to retrieve data, validate responses, or perform a state machine operation. An example workflow for a single question-answer pair is given in **Figure 5**, which is also used in Active Development #1 below. Based on information stored in this participant's electronic medical record (EMR), SmartState can retrieve relevant information based on a given question. In this case, the EMR indicates that this participant should take Acebutolol every morning. A researcher would like to know if this participant is still actively taking this medication. When prompted, the system asks, "When did you last take your beta blocker?" the user answers, "I took it this morning with my coffee." Using tool calling and the defined functions in the "Tool Chain," the system attempts to map the response to a function. In this case, the "Did Take Medication" function would be used to compare the response to the EMR data and verify that the response is adequate. Depending on the value produced from this function, the system either restates the question differently or continues to the next question. If the participant continues not answering the question appropriately after three attempts, a staff member is notified to contact the participant for an intervention.

Privacy is also a concern in any system that interacts directly with participants. To address this, the entire system operates within a NIST SP 800-53 and HIPAA-compliant datacenter at the University of Kentucky. Data collected during interactions is securely managed through several layers of protection. All data is encrypted both in transit and at rest, utilizing industry-standard encryption protocols to safeguard against unauthorized access. Access to the system is strictly controlled through the accompanying website, where researchers must log in using University credentials. Regular audits and updates to security protocols are conducted to address new threats and maintain the system's compliance.

In addition to handling participant responses, SmartState is designed to address potential server and network disruptions. To safeguard against data loss during such interruptions, SmartState automatically saves its internal state to stable storage every 15 minutes. By taking frequent snapshots, the system can restore all saved data (timed events, states of participants, scheduled messages, and conversations) and resume operations once the server or network issue is resolved.

**4. Results (Implementation)**

The development of SmartState was driven by the need to improve the efficiency and effectiveness of tasks related to interacting with study participants. In the first phase, a programmatic interface was provided for researchers to send and track participant text messages. Next, the study protocol was captured in a state graph, which automated data collection and interactions. This protocol automation has been in service for the last three years, greatly reducing the need for frequent human interaction while improving the participants' experience. The following section discusses an implementation of SmartState supporting an active research effort.

***Time-Restricted Eating Intervention in Postmenopausal Women***

Researchers at the University of Kentucky (UK) Department of Biology and Arizona State University College of Health Solutions are conducting a randomized clinical trial to assess the efficacy of time-restricted eating in decreasing metabolic risk. This trial requires collecting participants' calorie start and end times daily through text messaging. Depending on whether eating occurred during the correct time window, participants receive encouraging or informative feedback about their success rates. **Figure 6(a)** illustrates the simplified interaction of start and end calorie messages, which is essential for ensuring that participants adhere to the study protocol. This interaction requires participants to send "endcal" only after "startcal," which supports tracking eating windows consistently. **Figure 6(b)** shows the entire state machine, incorporating edge cases, reminders, and time-based conditions. This complexity highlights the necessity of automating these processes. At the beginning of the study, these messages were manually sent by the study team, as there were only a few participants enrolled. However, as participant enrollment increased, manual texting became infeasible. The system described in this paper was created to decrease the research team's workload and accommodate a larger number of participants. As the study evolved, the system expanded into multi-tenancy (many study protocols in one instance of SmartState) and grew to support 163 total participants. It is important to note that this number reflects only the participants recruited for the study and is not a limitation of SmartState's capabilities, which are designed to accommodate larger participant groups (see section **6.1 Performance Analysis** below for metrics).

Two metrics were evaluated in this study to measure the efficacy of the system: average success rate and average error rate. The success rate is defined as the average percentage of participants who successfully complete a fast. Successfully completing a fast means that a participant started and ended their calorie consumption within a window of nine to eleven hours. In contrast, an unsuccessful fast fell before 9 hours or after 11 hours. Each participant starts with a success rate of 100%. If a participant consumes calories for too long or too short, this success rate decreases. To determine the average of this rate, the total number of successful fasts is divided by the total number of days enrolled in the study. Based on all applicable participants, either currently or previously enrolled between September $9^{th}$, 2021, and September $12^{th}$, 2024, an average success rate of 92% was achieved.

The error rate is measured by the number of occurrences where the system does not understand a text message. For example, the most common error happens when a participant sends an ambiguous time, such as "startcal 7". It cannot be known if the participant means 7 AM or PM. Therefore, a response message will be sent saying, "Your STARTCAL time was not understood. Please send 'STARTCAL' again with your starting time, including 'am' or 'pm.'" This response message would be counted as an error. Other messages that would trigger similar responses include the addition of other words, phrases, or misspelling "startcal" or "endcal." From September $9^{th}$, 2021, to September $12^{th}$, 2024, 858 incoming messages needed additional context and were not recognized by the system. With the number of incoming messages totaling 24,403, the error rate was 3.5%.

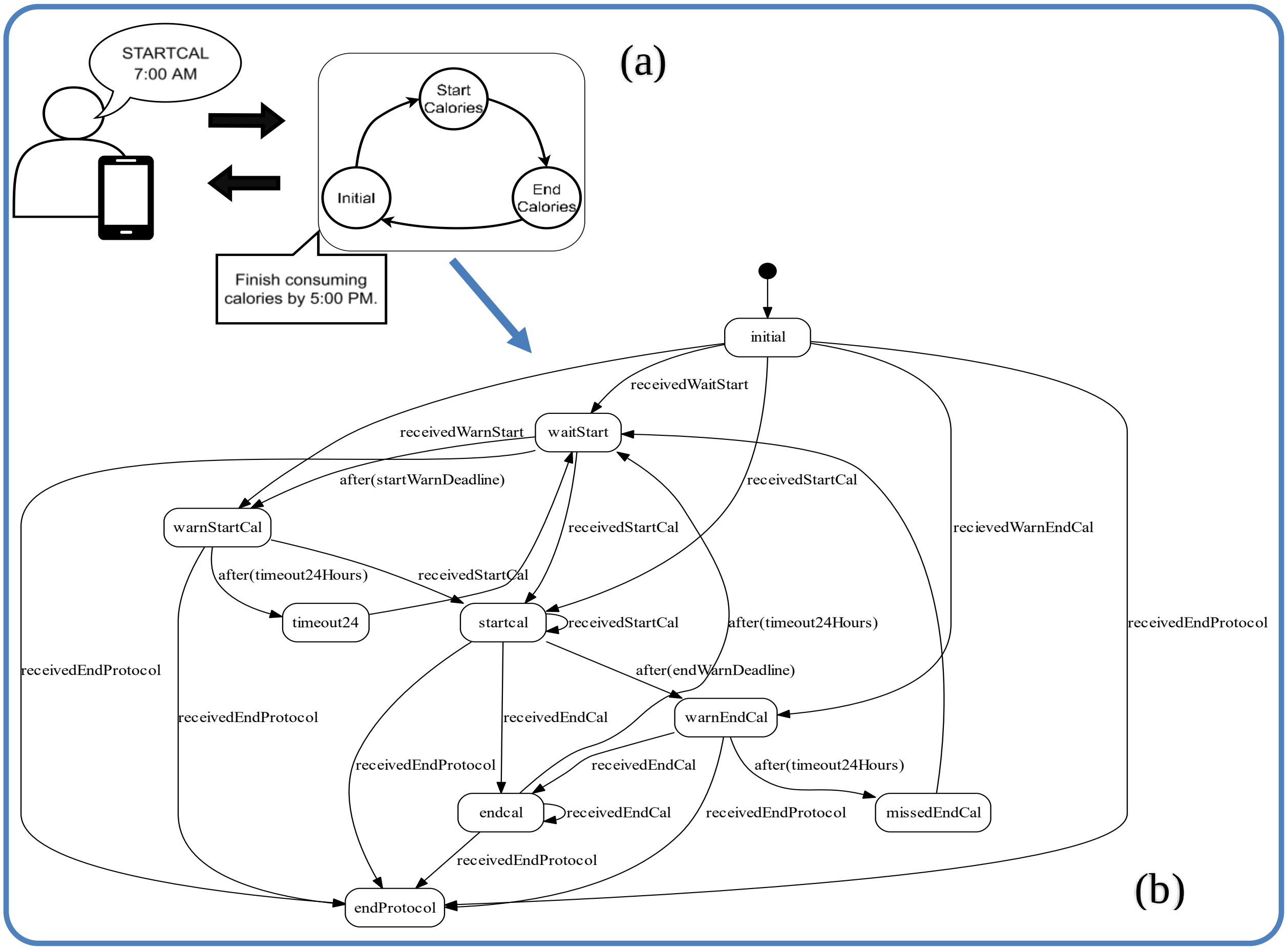

**Figure 6. (a)** Simplified state-based interaction for start and end calorie messages. **(b)** Full protocol state graph.

After implementing the system, researchers and participants no longer communicated directly about the start and end times for calorie tracking. Instead, the research team focused solely on registering new participants and analyzing the collected data. As of September 12th, 2024, the system has received 13,255 messages and sent 11,148 messages automatically without the need for intervention from research staff. In an informal survey of participants enrolled in the study before and after the implementation of SmartState, all respondents preferred the consistent reminders and near-instant responses produced automatically by the system.

## 5. Discussion

This discussion addresses SmartState's performance and future developments. **Section 5.1** highlights messaging and hardware limitations, with scaling solutions and a focus on data security. **Section 5.2** previews two research projects—one on preoperative care, the other on plant-based diets—demonstrating the system's flexibility in improving patient engagement and data collection across different studies.

### 5.1 Performance Analysis

When evaluating any software system, it is essential to identify and address potential bottlenecks. In this case, limitations primarily stem from the hardware supporting the system rather than the system itself. Two key areas that may experience slowdowns are the messaging service and LLM conversations.

First, to facilitate messaging between the system and participants, we utilize Twilio, a popular cloud communications platform that supports text messaging, calling, and email services. Twilio is known for its reliability and cost-efficiency, which is the reason we selected it for our messaging needs. However, the service enforces rate limits on message throughput, allowing up to 500 incoming[11] and 100 outgoing[12] messages per second for each 10-digit phone number. For higher message volumes, this limitation can be increased by adding additional phone numbers to a pool, raising the messaging capacity to scale dynamically with demand.

Second, the ability to process chat messages in real time depends heavily on the system's hardware. Currently, the SmartState system is hosted on a Microsoft Azure virtual machine with 2 CPU cores and 4GB of RAM, while the LLM is deployed on an Nvidia A100 Inference server. This setup allows for handling 40 participant messages per second. This limit was never reached with the implementation described above, as participant messages were received throughout the day rather than being concentrated on specific times. If this limit is reached, this capacity can be further improved by upgrading server hardware or utilizing alternative LLM services. For instance, commercial models like ChatGPT provide significantly higher throughput due to their vast infrastructure, though using these third-party services involves potential privacy concerns.

In contrast, using a local Llama instance and others for the implementations described in this paper ensures data security within a NIST- and HIPAA-compliant environment, which is crucial for handling sensitive participant information. All other components of SmartState are also contained in this environment. This local setup leverages the infrastructure of the UK's Institute for Biomedical Informatics' enterprise data center (EDC). High-performance inference servers, SQL Server databases, and an extensive Dell Isilon storage cluster are included here in a secure environment. The system's use of these machines balances performance and scalability with security and compliance.

### 5.2 Current/Future Efforts

As mentioned before, two additional research studies have chosen to implement SmartState. Both studies in development will be launched in the coming months, as of September 12th, 2024. We discuss these studies to demonstrate the flexibility of additional use cases for SmartState.

<u>Active Development #1</u>*: Optimization of Preoperative Treatment & Interactive Medical Assistance for Learning in Cardiothoracic Surgery (OptimalCT)*
OptimalCT is a collaboration between the UK Division of Cardiothoracic Surgery, the Society of Thoracic Surgery (STS), and the UK Center for Applied AI. This initiative focuses on improving adherence to preoperative protocols, such as beta-blocker administration, which is known to significantly reduce adverse outcomes in cardiac surgery[13]. National guidelines have been established to track and publicly report adherence to these protocols. However, UK's performance in documenting adherence, as reflected in the STS database, has been suboptimal, with a current rate of 78.3% compared to the national average of 94.0%. Since UK has a large number of outside referrals, this gap underscores the communication and preoperative optimization challenges. Despite these hurdles, SmartState can assist clinicians and patients in recording and reporting medication adherence.

There are three goals to address these issues. First, a patient must be established in the system during the outpatient clinic visit where the surgery is scheduled, medications are prescribed, and patient education occurs. Second, patient medication compliance must be monitored throughout the weeks before surgery. Finally, documentation of the administration of a beta blocker on the morning of surgery before arriving at the clinic must be confirmed and recorded in the STS database. Through timely, structured communications, SmartState communicates reminders and relevant information throughout the duration of the protocol. Specifically, the system reinforces medication instructions, using an LLM to retrieve and provide information about prescribed medications during patient conversations. Regular check-ins occur throughout the preoperative period to track adherence to protocols. Finally, on the day of surgery, SmartState prompts the patient to confirm beta blocker compliance, which is then recorded.

<u>Active Development #2</u>*: Plant-Based Diet for Diabetes Prevention*
In collaboration with the UK Athletic Training and Clinical Nutrition Department, this study is an open-label, non-randomized clinical trial with a 5-week, plant-based diet to analyze insulin sensitivity and skeletal muscle. All participants will follow the same diet, adjusted to meet their individual energy needs. The primary clinical outcome is the percentage change in the Matsuda insulin sensitivity index, calculated from OGTT insulin and glucose values. Another outcome will involve RNA-seq analyses of skeletal muscle, with comparisons made before and after the diet and between sexes. Pathway analysis will explore how diet-affected pathways differ by sex based on pre- and post-diet changes.

Previously, participants used a meal-logging app alongside manual text messages to ensure dietary compliance. The research team plans to replace the meal logging application and manual texting with SmartState. During the study, participants take photographs of their meals and snacks before and after eating. This is to verify they are eating the pre-approved meals and to estimate food intake. Participants will also rate each meal to inform the study team of which meals were disliked, potentially removing them from future studies. SmartState will also sends reminders to participants to capture meal images and prompt them if they are not submitted within a specific time. If participants miss the image submission, they will be asked to describe the meal via texting, ensuring no data is lost.

One of the planned feature developments for this study is the implementation of image analysis for participants' photos taken before and after their meals. By leveraging techniques such as image segmentation and deep learning, we are confident that integrating this feature into SmartState will reduce the need for manual data review.

## 6. Conclusions

We have introduced a system for tracking and verifying participant protocol adherence throughout a time-restricted eating study, with additional studies being developed. SmartState eliminates the need for manual data collection while maintaining an auditable log of participants and system actions at each step over the course of a study.

By utilizing state-based architecture, the prevention of participants arbitrarily transitioning between states will ensure study integrity and consistency in its operation. This approach also reduces the workload for developers, enables automatic calculation of participant metrics, decreases message response time, and simplifies participant management for researchers. By abstracting complex studies into state graphs, FSM implementations can be quickly created to reduce development lead times and prevent unintentional bugs. We built on this FSM base by adding a messaging system to send and receive participant messages and an LLM to handle conversational tasks such as response adequacy. The implementation of this automated messaging system outperforms manual methods by increasing the consistency of scheduled messages and responding nearly instantly to participants.

Additionally, we offer a web-based platform for researchers to manage participants. This platform allows them to access audit and message logs and export participant data for analysis. It provides user-friendly interactions with the FSM, enabling researchers to add participants, modify study assignments, and monitor both current and past participants' states. Researchers can also manually transition participants between states if necessary.

The system's flexibility allows researchers to easily implement their own study requirements according to their specifications. The time-restricted eating study, along with other active developments described, shows how SmartState can be applied to automatically manage participant interactions in an efficient, verifiable, and robust way.

The open-source repository for SmartState is available here: https://github.com/innovationcore/SmartState-public.

**Acknowledgement**

This research was supported by the National Institute of Diabetes and Digestive and Kidney Diseases and the National Center for Advancing Translational Sciences of the National Institutes of Health under award numbers R01DK124774 and UL1TR001998. The content is solely the responsibility of the authors and does not necessarily represent the official views of the NIH.